\documentclass{aa}

\usepackage{txfonts}

\usepackage{graphicx}

\usepackage{natbib}
\bibpunct{(}{)}{;}{a}{}{,} 

\newcommand{\ten}[2]{#1\times 10^{#2}}

\newcommand{\rsun}{$\mathrm{R}_\odot$}
\newcommand{\msun}{$\mathrm{M}_\odot$}

\newcommand{\ori}{\mbox{\object{V1309\,Ori}}}
\newcommand{\rosat}{\textit{ROSAT}}

\newcommand{\msunyr}{$\mathrm{M}_\odot\mathrm{yr}^{-1}$}

\newcommand{\cm}{cm$^{-2}$}
\newcommand{\cmcub}{cm$^{-3}$}

\newcommand{\ergs}{erg\,cm$^{-2}$s$^{-1}$}
\newcommand{\ergsa}{erg\,cm$^{-2}$s$^{-1}$\AA$^{-1}$}

\begin{document}

\title{The secondary star and distance of the polar V1309~Ori\thanks{Based on observations obtained under run ID 01BK6 at the
 Canada-France-Hawaii Telescope (CFHT) which is operated by the
 National Research Council of Canada, the Institut National des
 Sciences de l'Univers of the Centre National de la Recherche
 Scientifique of France, and the University of Hawaii.}}
  \author{K.~Reinsch\inst{1} \and Y.~Kim\inst{2} \and K.~Beuermann\inst{1}}

  \institute{Institut f\"ur Astrophysik, Friedrich-Hund-Platz~1,
    D-37077~G\"ottingen, Germany,
    \email{reinsch@astro.physik.uni-goettingen.de} \and Dept. of
    Astronomy and Space Science, Chungbuk National University,
    Cheongju, 360-763 Korea
    \email{ykkim153@chungbuk.ac.kr}}

  \date{Received May 23, 2006\ / Accepted July 03, 2006}
  
\abstract { 
The first phase-resolved JHK light curves of the
eclipsing polar (AM~Herculis binary) \ori\ are presented and
interpreted.}  
{ 
We separate the contributions from the
secondary star and from other sources with the aim of determining a
photometric distance.}  
{ 
Simple model calculations show
that the accretion stream and the cyclotron source on the accreting
white dwarf are minor contributors to the infrared light, allowing an
accurate determination of spectral type and absolute flux of the
secondary star.}  
{ 
The unilluminated backside of the secondary star as seen in eclipse
has spectral type dM0 to dM0+. Its dereddened magnitude is $K=13.58$
at orbital phase $\phi=0$ (eclipse). Using the calibrated surface
brightness of M-stars and the published mass of the secondary,
$M_2=0.46$\,\msun, we obtain a distance $d=600\pm 25$\,pc which scales
as $M_2^\mathrm{1/2}$. The radius of the Roche-lobe filling secondary
exceeds the main-sequence radius of an M0 star by
$21^{+11}_{~-6}$\%.}
{ 
The debated origin of the infrared light of \ori\ has been settled in
favor of the secondary star as the main contributor and an accurate
distance has been derived that will place estimates of the luminosity and 
synchronization time scale on a more secure basis.}  
\keywords {Stars: binaries: eclipsing -- 
Stars: cataclysmic variables -- 
Stars: distances -- 
Stars: late-type --
Stars: magnetic fields --
Stars: individual (\ori)}
  
\titlerunning{The distance of \ori}
\authorrunning{K.~Reinsch et al.}

\maketitle


\section{Introduction}

The soft X-ray source RX\,J0515.6+0105 was discovered in the \rosat\
All-Sky-Survey \citep{beuermannthomas93}, quickly identified as an
eclipsing polar (or AM\,Herculis binary)
\citep{garnavichetal94,shafteretal95,walteretal95}, and named
\ori. It turned out to be a key object for studies of the
synchronization of polars because of its long orbital period of
7.98\,h, twice as long as that of any other polar, and the implied
larger separation of the components. At the same time, the magnetic
field strength of about 60\,MG \citep{garnavichetal94,shafteretal95}
and the accretion rate of $\ten{(1.0\pm 0.5)}{-9}$\,\msunyr\
\citep{beuermannburwitz95,harropallinetal97,schwarzetal05} are within
the ranges typical of long-period polars. Luminosity-based estimates
of the accretion rate depend sensitively on the bolometric
correction, the mass of the primary, and on the assumed
distance. Published distances of \ori\ differ considerably, ranging
from about 500 pc \citep{garnavichetal94,staudeetal01} to greater than
1500\,pc \citep{harropallinetal97}. The synchronization of a polar
requires that the magnetostatic torque between the assumed magnetic dipole
moments of the two stars in the system overcomes the accretion torque
on the white dwarf \citep{kingetal94,wuetal94}, with the latter
depending directly on the accretion rate. The parameters derived so
far place \ori\ near the limit of being synchronized and measuring its
luminosity and mass transfer rate is important for understanding the
physics of polars in general.

In this paper, we derive an accurate distance from the brightness of
the secondary in \ori. This result is based on our previous
calibration of the surface brightness of M-dwarfs in standard and
non-standard photometric bands \citep{beuermann00} and will place
estimates of the luminosity, accretion rate, and synchronization time
scale on a more secure basis.

\section{Observations and Data Analysis}

\ori\ was observed with the CFHT-IR camera at the CFHT 3.6-m telescope
on Mauna Kea, Hawaii, for 6.7 hours on 26/27 October 2001. During the
observations, the star was in its normal bright state. The filters
employed were J, H, K', and K. Almost complete phase coverage was
obtained with a gap only near orbital phase $\phi=0.4$, where $\phi=0$
indicates the center of the eclipse for the ephemeris of
\citet{staudeetal01}. Five 15 sec exposures taken in dithering
mode in a single filter were averaged to yield an individual data
point. Exposures in J, H, and K were obtained consecutively and the
sequence repeated in turn, leading to a time resolution of 13
min. Variability on shorter time scales is lost and only in this sense
are the observations in the three bands quasi-simultaneous and not
truly simultaneous. UKIRT faint standards FS\,150, 152, 153, and
SA\,92-342 were observed as standard stars and the results reduced
into the standard Mauna Kea infrared system
\citep{tokunagavacca05}. We interprete these data in combination with
our earlier BVRI photometry and low resolution optical
spectrophotometry \citep{shafteretal95} and the photometry of
\citet{katajainenetal03}.

\section{Light curves}

\begin{figure}[t]
\begin{center}
\includegraphics[width=8.6cm]{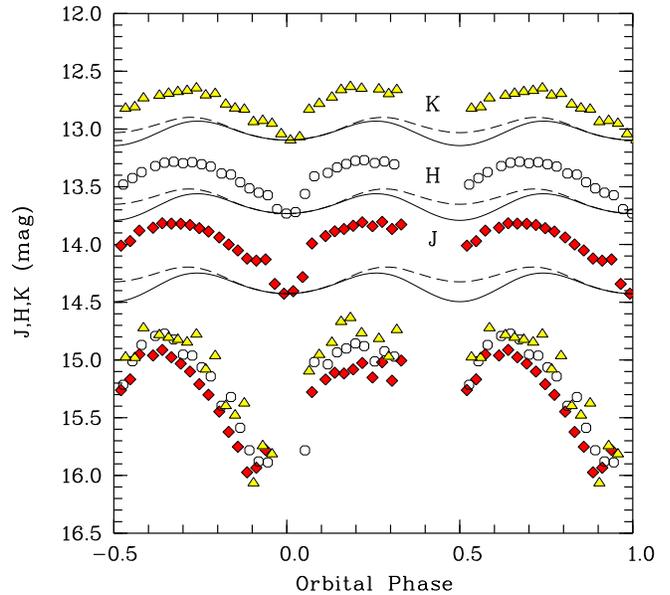}
\end{center}
\caption{Quasi-simultaneous near-infrared light curves of \ori\ in the
J, H, and K bands folded with the binary orbital period.  Also shown
are the calculated light curves of the ellipsoidal modulation of the
secondary added. The K-band light curve has been shifted upwards by
0.5\,mag in order to avoid overlap. The solid (dashed) curves refer to
the unirradiated (irradiated) secondary star. The data sets at the
bottom of the figure show the difference, representing infrared light
from sources other than the secondary (see text).}
\label{fig:lc}
\end{figure}

Figure~\ref{fig:lc} shows the light curves of \ori\ in the bands J, H,
and K folded over the orbital period of 7.98\,h using the ephemeris of
\citet{staudeetal01}. They display a quasi-sinusoidal modulation with
the eclipse superimposed. The depth of the eclipse is 0.31\,mag,
0.20\,mag, and 0.15\,mag in J, H, and K, respectively, much less than
in the optical where it reaches 2.1\,mag in B and nearly 4 mag in U
\citep{shafteretal95,katajainenetal03}. The light outside eclipse
arises from the accretion stream and and the cyclotron source in the
accretion spot on the white dwarf. This combined source evidently
still contributes in the infrared, although much less than in the
optical. Given our low time resolution, we can not measure ingress and
egress times of the eclipsed source.  Our data, however, are
consistent with the eclipse being total for a time interval similar to
that observed in the optical ($\sim40$ min).

At first glance, the double-humped structure of the light curve seems
to suggest that we see the ellipsoidal light modulation of the
secondary star. However, that is refuted by the phasing of the primary
eclipse and the finite eclipse depth. Many polars (and non-magnetic
CVs) contain light sources which mimic the ellipsoidal modulation. In
the case of polars, this can be the accretion stream which contributes
maximum flux when seen sideways and reaches a minimum when the plane
of the accreting field line crosses the line of sight. In \ori, this
is the case at $\phi\simeq0.91$. A double-humped light source can also
originate from cyclotron beaming and the varying flux level as the
angle between the accreting field line at the white dwarf and the line
of sight varies.  We find that the maxima at $\phi=0.2$ and $\phi=0.7$
are of nearly equal height in the infrared, while the $\phi=0.2$
maximum is the dominant one in the optical. The available results
combined suggest that the light curves are shaped by the varying
geometrical aspect, mutual occultation of the stream by itself or the
stream and the secondary star, and wavelength-dependent illumination
by the soft X-ray and cyclotron sources on the white dwarf. We refrain
from modeling the light curves in detail.

We wish to derive the spectral energy distributions (SED) inside and
outside eclipse and, for this purpose, we have to model the
ellipsoidal light modulation of the secondary star. We assume that the
eclipse is total \citep{schmidtstockman01,staudeetal01} and that the
flux observed at eclipse center arises from the backside hemisphere of
the secondary, i.e. the one opposite to the L$_1$ point. We calculate
the ellipsoidal light curves using
\textit{Nightfall}\footnote{\hspace*{-1.5mm}http://www.hs.uni-hamburg.de/DE/Ins/Per/Wichmann/Nightfall.html}. This
program uses the Roche geometry, the limb darkening coefficients of
\citet{claret98}, and the gravity darkening coefficients of
\citet{claret00} that describe the transition from high-temperature
radiative stars to low-temperature convective stars. It allows the use
of Hauschildt's model atmospheres for non-irradiated stars, but for a
simple application including irradiation one fares better with the
blackbody approximation. We have adjusted the resulting ellipsoidal
light curves at $\phi=0$ and show them both, for the case of vanishing
irradiation (solid curves) and for irradiation by a source at the
location of the white dwarf with a luminosity of $10^{34}$\,\ergs\
(dashed curves). This luminosity corresponds to the rather high
accretion rate of $\ten{1.5}{-9}$\,\msunyr. The influence of
irradiation is seen to be moderate in the infrared, while the minimum
at $\phi=0.5$ nearly vanishes in the optical for the parameters chosen
here. We obtain the SEDs at $\phi=0$ and at $\phi=0.21$ by combining
our infrared photometry with the our earlier BVRI photometry, our
3600-9140\AA\ spectrophotometry \citep{shafteretal95}, and the
photometry of \citet{katajainenetal03}.

\begin{table}[b]
\caption{Dereddened magnitudes of the secondary star in \ori\ at
orbital phase $\phi=0$ (center of eclipse) compared with the
magnitudes of M0 and M1 field adjusted to the same mean flux level as
\ori.}
\label{tab:sec}
\centering
\begin{tabular}{lccc} 
\hline \hline \noalign{\smallskip}
Band           & \ori\ &  M0   &  M1   \\ 
\noalign{\smallskip} \hline
\noalign{\smallskip}
$V$            & 17.22 & 17.17 & 17.40 \\
$I_\mathrm{c}$ & 15.42 & 15.43 & 15.48 \\
$J$            & 14.39 & 14.37 & 14.28 \\
$H$            & 13.71 & 13.76 & 13.67 \\
$K$            & 13.58 & 13.61 & 13.49 \\
\noalign{\smallskip} \hline      
			         
\end{tabular}
\end{table}

\section{Spectral energy distributions}

\begin{figure}[t]
\begin{center}
\includegraphics[width=8.6cm]{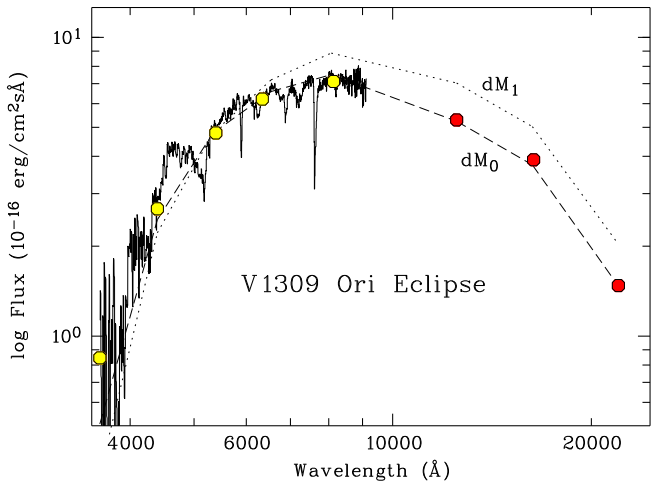}
\end{center}
\caption{The spectral energy distribution of the remnant light in the
central phase intervals of the eclipse is consistent with an origin
from the secondary in \ori\ alone. Besides our JHK photometry (filled
circles), published UBVRI photometry (open circles) and the
\citet{shafteretal95} spectrophotometry (solid line) are
shown. Included for comparison is the flux distribution of a
young-disk M0 dwarf (dashed line) adjusted to the mean flux level of
\ori\ and that of an M1 dwarf adjusted at V (dotted line).}
\label{fig:sec}
\end{figure}

Figure~\ref{fig:sec} shows the SED at $\phi=0$, dereddened with
$E_\mathrm{B-V}=0.15$ \citep{schmidtstockman01}. The Shafter et
al. spectrophotometry in eclipse is slightly adjusted by a quadratic
function in wavelength to fit the BVRI photometry from the same
reference. The data point for the U-band is from
\citet{katajainenetal03}. The SED is characterized by a color
$V-K=3.64$ (see Tab.~\ref{tab:sec}). This value closely resembles that
of a dwarf of spectral type M0 to M0+. The M0 SED is shown as a dashed
line in Fig.~\ref{fig:sec} that is adjusted to the observed VIJHK data
points and is seen to fit extremely well. The SED of an M1 dwarf
(dotted line) is clearly too red. In the U-band, an excess of the
photometry over the M0 SED may exist and possibly also a slight excess
in B. A possible cause is an atmospheric structure of the secondary
star in \ori\ which differs slightly from that of a field star, even
on the non-irradiated hemisphere. Alternatively, a 2\% leak from the
bright phase spectrum would also explain the UB excess, but the weak
emission lines expected in that case are not visible in
eclipse. Furthermore, the Hubble Space Telescope phase-resolved
ultraviolet spectroscopy \citep{schmidtstockman01,staudeetal01} and
the model calculations of \citet{katajainenetal03} suggest that the
eclipse of the accretion stream is total. No deviation from the SED of
an M0 to M0+ star beyond the $\sim0.03$ mag scatter is seen for all
bands from V to K. The excellent agreement confirms that the infrared
flux observed in eclipse represents the secondary star. 

\begin{figure}[t]
\begin{center}
\includegraphics[width=8.6cm]{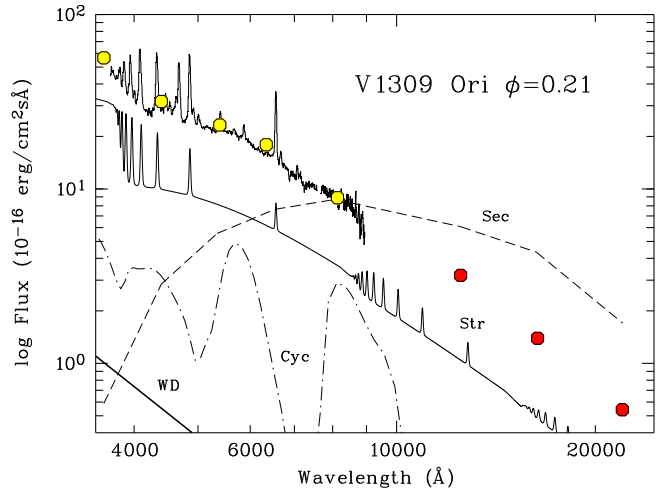}
\end{center}
\caption{Spectral energy distribution of the light from sources other
than the secondary at orbital phase $\phi =0.21$. The symbols are as
in Fig.~2. The (subtracted) contribution of the secondary at $\phi
=0.21$ is shown as the dashed line. A spectrum representative of the
hydrogen emission of the stream (solid line) and a cyclotron spectrum
from the accretion spot (dot-dashed line) are shown for
comparison. Their normalization is arbitrary (see text). }
\label{fig:stream}
\end{figure}

The SED for $\phi=0.21$, with the contribution from the secondary star
(dashed line labeled `Sec') subtracted, is shown in
Fig.~\ref{fig:stream}. The dereddened Shafter et al. spectrophotometry
is shown but not adjusted to the photometry in this case. Although the
optical and infrared observations are not simultaneous, an internally
consistent distribution emerges which falls off with wavelength
approximately as $f_\lambda \propto \lambda^{-2.6}$. The secondary
star is obviously the dominant source in the infrared, while the other
sources dominate in UBVR and both contribute equally in the I band.
The secondary contributes 3/4 of the observed flux in K which clearly
contradicts the claim by \citet{harropallinetal97} of a contribution
of only 20\%. These authors based their statement on the lack of
detectable spectroscopic M-star features in their infrared
spectrum. Although we have no spectral resolution in the infrared, we
consider our conclusion on the SED of the secondary star firm and
suggest instead that the Harrop-Allin et al. infrared spectrum, which
was taken around $\phi=0.5$, is affected by irradiation of the
secondary star and may differ from that of a field star.

Although we do not embark on a complete modeling of the $\phi=0.21$
SED, we comment semi-quantitatively on the two main contributing light
sources. The Shafter et al. spectrophotometry at this phase shows the
Balmer and the Paschen jumps in emission along with the appropriate
emission lines suggesting a temperature around 10000--20000\,K. Our
heuristic model consists of two main contributions to the observed
SED. As solid curve labeled `Str', we show the emission from an
isothermal hydrogen plasma at a temperature of 15000\,K with two
subcomponents of different density and geometrical extent: the first
subcomponent has a particle density of $\ten{5}{13}$\,\cmcub, a depth
of $10^9$\,\cm, unity relative projected area, and represents the
outer accretion stream; the second subcomponent has a particle density
of $\ten{3}{14}$\,\cmcub, a depth of $\ten{3.5}{8}$\,\cm, a relative
projected area of 0.35, and represents the inner accretion stream a
factor of two in radius closer to the white dwarf. Stark broadening is
included and the resolution of the resulting spectrum adjusted to the
35\AA\ FWHM resolution of the spectrophotometry. Only the sum of the
two subcomponents is shown with an arbitrary normalization. This
spectrum fits the shape of the observed optical spectrum reasonably
well and is seen to fall off in the infrared with a similar slope 
as the observed
SED. As the dot-dashed curve labeled `Cyc', we show the cyclotron
emission from an isothermal plasma with a temperature of 5\,keV in a
magnetic field of 63\,MG for a combination of two depth parameters,
log\,$\Lambda=6$ which yields a largely optically thick spectrum, and
log\,$\Lambda=2$ which accounts for the optically thin emission
lines. The latter fits the two faint and broad bumps at about 5800\AA\
and 8500\AA\ reported by \citet{shafteretal95} that represent the
second and third cyclotron harmonics in a field of 63\,MG. Our model
spectrum demonstrates that the fourth harmonic near 4300\AA\ is
already weak in the log\,$\Lambda=2$ spectrum, while the first
harmonic at 1.7$\mu$m is largely optically thick (and off the
scale). Hence, we do not expect to observe more than the two detected
cyclotron harmonics. The high density of the accretion stream at the
white dwarf deposits much of the accretion energy at sub-photospheric
levels \citep{walteretal95, schwarzetal05} and explains why cyclotron
line emission and polarization are weak
\citep{garnavichetal94,shafteretal95,buckleyshafter95}.  The observed
overall SED is quite well reproduced by a linear combination $f_\lambda
= 2.5\,f_\mathrm{str}+0.5\,f_\mathrm{cyc}$. The white dwarf represents
a minor contribution as shown by the thick line labeled `WD' which is
based on the upper limit to the white dwarf contribution as derived
by \citet{staudeetal01}.
A quantitative model could be constructed along these lines, but the
optical-depth effects are substantial and one may expect a sizeable
fraction of the stream emission to fall in the EUV (Lyman jump in
emission), where no observations exist. It is doubtful, therefore,
whether convincing results for the bolometric luminosity and the
accretion rate can be obtained this way; we do not attempt it here.

\section{Distance of \ori}

We determine the distance to \ori\ by the surface brightness method
\citep{barnesevans76,bailey81,ramseyer94,beuermannetal99,beuermann00}
which connects the distance $d$, the stellar radius $R$, the surface
brightness $F_\lambda$ at wavelength $\lambda$, and the dereddened
observed flux $f_\lambda$ by the inverse square law
\begin{equation}
F_\lambda  = f_\lambda\left(\frac{d}{R}\right)^2 .
\label{eq:square}
\end{equation}
This relation is often quoted with the surface brightness and the flux
expressed in magnitudes ($S_\lambda$ and $m_{\lambda}$) as
\begin{equation} 
{\rm log}\,d = (m_{\lambda} - S_{\lambda})/5 + 1 + {\rm
log}(R/R_{\odot}).
\label{eq:dist}
\end{equation} 
Using a sample of main sequence stars with appropriate metalicity,
the $F_\lambda$ or $S_\lambda$ for different photometric bands have
been calibrated as functions of color or spectral type
\citep[][]{beuermannetal99,beuermann00,beuermann06}. The distance
is found to be proportional to the stellar radius of the secondary
star which is given by Roche geometry.

\citet{staudeetal01} quote the mass range of the secondary in \ori\ as
0.4--0.6\,\msun, with a most probable value of
$M_2=0.46$\,\msun.  The radius of the Roche-lobe filling secondary
scales as $M_2^{1/3}$ and is nearly independent of the white dwarf
mass. The appropriate projected area is that of the Roche lobe as seen
at $\phi=0$ at an inclination of $78^\circ$ \citep{staudeetal01}. The
corresponding effective mean radius is
$R=\ten{4.88}{10}\,\mathrm{cm}=0.701$\,\rsun.  We use
Eq.~(\ref{eq:dist}) with $K=13.58$ and $S_\mathrm{K}=3.91$ for
$V-K=3.64$ \citep{beuermann00} to obtain $d=602$\,pc. If there is some
blue excess in the SED of the secondary it might be more appropriate
to use $V-K$ larger by 0.10 and $S_\mathrm{K}=3.94$ and which yields
$d=594$\,pc. The statistical and systematic errors in the measurement
of $K$ and the calibration of $S_\mathrm{K}$ of about 0.05 each lead
to a 4\% error in $d$. We settle on $d=600\pm25\,^{+55}_{-27}$\,pc,
where the second error refers to the full range of $M_2$ from
\citet{staudeetal01}. Instead of $K$ and $S_\mathrm{K}$, we might have
used any other photometric band from $V$ to $H$. The results would be
similar because the secondary corresponds so closely to the SED of an
M0 to M0+ field star. The error, however, is smallest in $K$ because
$S_\mathrm{K}(V-K)$ has the largest lever arm and the shallowest slope
of all surface brightness vs. color relations. As an additional
example, we quote the result for the narrow-band surface brightness
$F_{7500}$ at 7500\AA\ which was also calibrated by
\citet{beuermann00}. From the SED in Fig.~\ref{fig:sec} one finds a
dereddened flux $f_{7500}=\ten{7.19}{-16}$\,\ergsa. With
$F_{7500}=\ten{(1.16\pm0.07)}{6}$\,\ergsa\ \citep{beuermann00} for a
star of spectral type M0 to M0+, Eq.~(\ref{eq:square}) yields
$d=635$\,pc with a slightly larger error than above\footnote{The
  distance of 745\,pc quoted by Beuermann (2000) did not include
  dereddening and was based on an a slightly different calibration of
  $F_{7500}$.}.

Our result is in excellent agreement with the distance obtained from
the comparison of the brightness of the secondary star in \ori\ with
that of the mean component of the binary YY~Gem which yields 625\,pc
\citep[][corrected to the Hipparcos distance of YY~Gem by Schwarz et
al. 2005]{staudeetal01}. The much larger distance suggested by
\citet{harropallinetal97}, $d > 1500$\,pc, based on the assignment of
80\% of the $K$-band flux to the accretion stream, can be excluded.

The distance modulus to \ori\ is $K-M_\mathrm{K}=
8.89^{+0.19}_{-0.10}$ and with a $4\pi$ average magnitude of the
unheated secondary of $K=13.48$, its absolute $K$-band magnitude would
be $M_\mathrm{K}=4.59^{+0.10}_{-0.19}$, where the errors refer to the
full range of secondary masses of 0.4 to 0.6\,\msun, with 0.46\,\msun\
the nominal mass. A dM0+ main sequence star of near-solar metalicity
has $M_\mathrm{K}=5.0$ and a radius $R$ given by
log($R$/\rsun)\,=\,--0.22 \citep{beuermannetal98,beuermannetal99},
while the volume radius of the Roche-lobe filling secondary in \ori\
is $R=\ten{(5.04^{+0.47}_{-0.23})}{10}$\,cm with
log($R$/\rsun)\,=\,$-0.14^{+0.04}_{-0.02}$. Hence, we find that the
Roche-lobe filling secondary is expanded over a main sequence star of
the same spectral type by $(21^{+11}_{~-6})$\%, as noted first by
\citep{garnavichetal94}. This oversize may indicate moderate nuclear
evolution. On the other hand, the secondary in V471~Tau which does not
fill its Roche lobe has a radius 18\% larger than a main sequence
K-star, probably because a substantial fraction of its surface is
covered by star spots \citep{obrienetal01}. Activity may, therefore,
be an alternative cause for the observed inflation of the secondary in
\ori. A measurement of the metal abundances supplemented by
evolutionary modeling may possibly distinguish between the two
possibilities.

\section{Conclusions}

We have presented the first phase-resolved infrared photometry of
\ori\ and deduced the spectral energy distributions (SED) of the
secondary star and of the additional light sources outside eclipse for
the 3600--22000\,\AA\ range. The secondary is of spectral type M0 to M0+
and contributes 75\% of the light in the $K$-band. We derive a
distance of 600\,pc from its $K$-band magnitude. The SED at orbital
phase $\phi=0.21$ can be plausibly explained as the superposition of
emission from the accretion stream and the cyclotron source on the
white dwarf.

\begin{acknowledgements}
This research was supported by the Korea Astronomy Observatory and 
Space Science Institute Research Fund 2002 and by the Deutsches Zentrum
f\"ur Luft- und Raumfahrt e.V. (DLR) under grant 50\,OR\,99\,03\,6. 
We are deeply indebted to Y.-B. Jeon for generous support of our 
observing proposal within the KAO-CFHT collaboration. We thank Boris
T. G\"ansicke for providing his program to calculate the emission
from a hydrogen plasma and Frederic V. Hessman for comments on the
manuscript.
\end{acknowledgements}

\bibliographystyle{aa}

\begin{thebibliography}{29}
\expandafter\ifx\csname natexlab\endcsname\relax\def\natexlab#1{#1}\fi

\bibitem[Bailey (1981)]{bailey81} Bailey, J.\ 1981, \mnras, 197, 31

\bibitem[Barnes \& Evans (1976)]{barnesevans76} Barnes T.G., Evans
D.S.\ 1976, \mnras, 174, 489

\bibitem[Beuermann (2000)]{beuermann00} Beuermann, K.\ 2000, New Astr. Rev., 44, 93

\bibitem[Beuermann (2006)]{beuermann06} Beuermann, K.\ 2006, \aap, submitted

\bibitem[Beuermann et al. (1998)]{beuermannetal98} Beuermann K.,
Baraffe I., Kolb, U., Weichhold, M.\ 1998, \aap, 339, 518

\bibitem[Beuermann et al. (1999)]{beuermannetal99} Beuermann K.,
Baraffe I., Hauschildt, P.\ 1999, \aap, 348, 524

\bibitem[Beuermann \& Burwitz (1995)]{beuermannburwitz95} Beuermann,
K., \& Burwitz, V.\ 1995, ASP Conf. Ser., 85, 99

\bibitem[Beuermann \& Thomas (1993)]{beuermannthomas93} Beuermann, K.,
\& Thomas, H.~C.\ 1993, Adv. Space Res. 13(12), 115

\bibitem[Buckley \& Shafter (1995)]{buckleyshafter95} Buckley,
D.~A.~H., \& Shafter, A.~W.\ 1995, \mnras, 275, L61

\bibitem[O'Brien et al.(2001)]{obrienetal01} O'Brien, M.~S., Bond, 
H.~E., \& Sion, E.~M.\ 2001, \apj, 563, 971 

\bibitem[Claret (1998)]{claret98} Claret, A.\ 1998, \aaps, 131, 395

\bibitem[Claret (2000)]{claret00} Claret, A.\ 2000, \aap, 359, 289

\bibitem[Garnavich et al. (1994)]{garnavichetal94} Garnavich, P.~M.,
Szkody, P., Robb, R.~M., Zurek, D.~R., \& Hoard, D.~W.\ 1994, \apj, 435, L141

\bibitem[Harrop-Allin et al. (1997)]{harropallinetal97} Harrop-Allin,
M.~K., Cropper, M., Potter, S.~B., Dhillon, V.~S., \& Howell, S.~B.\
1997, \mnras, 288, 1033

\bibitem[Katajainen et al. (2003)]{katajainenetal03} Katajainen, S.,
Piirola, V., Ramsay, G., et al.\ 2003, \mnras, 340, 1

\bibitem[King et al. (1994)]{kingetal94} King, A.~R., Kolb, U., de
Kool, M., \& Ritter, H.\ 1994, \mnras, 269, 907

\bibitem[Ramseyer (1994)]{ramseyer94} Ramseyer, T.~F.\ 1994, \apj, 425, 243

\bibitem[Schmidt \& Stockman (2001)]{schmidtstockman01} Schmidt,
G.~D., \& Stockman, H.~S.\ 2001, \apj, 548, 410

\bibitem[Schwarz et al. (2005)]{schwarzetal05} Schwarz, R., Reinsch,
K., Beuermann, K., \& Burwitz, V.\ 2005, \aap, 442, 271

\bibitem[Shafter et al. (1995)]{shafteretal95} Shafter, A.~W.,
Reinsch, K., Beuermann, K. et al.\ 1995, \apj, 443, 319

\bibitem[Staude et al. (2001)]{staudeetal01} Staude, A., Schwope,
A.~D., \& Schwarz, R.\ 2005, \aap, 374, 588

\bibitem[Tokunaga  \& Vacca (2005)]{tokunagavacca05}  Tokunaga, A.~T.,
\& Vacca, W.~D.\ 2005, \pasp, 117, 421

\bibitem[Walter et al. (1995)]{walteretal95} Walter, F.~M., Wolk,
S.~J., \& Adams, N.~R.\ 1995, \apj, 440, 834

\bibitem[Wu et al. (1994)]{wuetal94} Wu, K., Wickramasinghe, D.~T.,
Bailey, J., Tennant, A.~F.\ 1994, PASA, 11, 198
 
\end{thebibliography}

\end{document}